\documentclass[useAMS,usenatbib]{mnras}
\usepackage{savesym}
\usepackage{txfonts}
\savesymbol{iint}
\savesymbol{iiint}
\savesymbol{iiiint}
\savesymbol{idotsint}
\usepackage{graphicx}
\usepackage{amssymb}
\usepackage[below]{placeins}
\usepackage{txfonts}
\usepackage{natbib}
\usepackage{amsmath}
\usepackage{float}
\restoresymbol{TXF}{iint}
\restoresymbol{TXF}{iiint}
\restoresymbol{TXF}{iiiint}
\restoresymbol{TXF}{idotsint}
\usepackage{multirow}
\usepackage{array}
\usepackage{xcolor}
\usepackage[percent]{overpic}
\bibpunct{(}{)}{;}{a}{}{,}

\def \aap{A\&A}
\def \apjl{ApJ}

\def \mnras{MNRAS}
\def \apj{ApJ}

\def\nat{Nature}

\def\spose#1{\hbox to 0pt{#1\hss}}
\def\approxlt{\mathrel{\spose{\lower 3pt\hbox{$\sim$}}
        \raise 2.0pt\hbox{$<$}}}
\def\approxgt{\mathrel{\spose{\lower 3pt\hbox{$\sim$}}
        \raise 2.0pt\hbox{$>$}}}
\def\approxpropto{\mathrel{\spose{\lower 3pt\hbox{$\sim$}}
        \raise 2.0pt\hbox{$\propto$}}}
\mathchardef\twiddle="2218

\def\multleft#1{\hbox to size{\vbox {\halign {\lft{##}\cr #1}}\hfill}\par}
\def\multright#1{\hbox to size{\vbox {\halign {\rt{##}\cr #1}}\hfill}\par}

\def\today{\ifcase\month\or January\or February\or March\or April\or May\or
      June\or July\or August\or September\or October\or November\or December\fi
      \space\number\day, \number\year}
\def\<{\thinspace}

\def\arcsec{{\rm\thinspace arcsec}}

\def\arcsec {\hbox{$^{\prime\prime}$}}

\makeatletter
\newcommand{\thickhline}{%
    \noalign {\ifnum 0=`}\fi \hrule height 1.2pt
    \futurelet \reserved@a \@xhline
}
\newcolumntype{"}{@{\hskip\tabcolsep\vrule width 1pt\hskip\tabcolsep}}
\makeatother

\bibpunct{(}{)}{;}{a}{}{,}

\title[ALMA view of molecular gas in M87]{ALMA observation of the disruption of molecular gas in M87}

\author[Simionescu et al.]
{\parbox{\textwidth}{A. Simionescu$^{1}$, G. Tremblay$^{2,3}$, N. Werner$^{4,5,6}$, R. E. A. Canning$^{7,8}$, S. W. Allen$^{7,8,9}$, J. B. R. Oonk$^{10,11}$ }\
\vspace{0.5cm}\\
\parbox{\textwidth}{$^1$Institute of Space and Astronautical Science (ISAS), JAXA, 3-1-1 Yoshinodai, Chuo-ku, Sagamihara, Kanagawa, 252-5210, Japan\\
$^2$Department of Physics and Yale Center for Astronomy \& Astrophysics, Yale University, 217 Prospect Street, New Haven, CT 06511, USA \\
$^{3}$Harvard-Smithsonian Center for Astrophysics, 60 Garden St, Cambridge, MA 02138, USA\\
$^{4}$MTA-E\"otv\"os Lor\'and University Lend\"ulet Hot Universe Research Group, H-1117 P\'azm\'any P\'eter s\'et\'any 1/A, Budapest, Hungary\\
$^{5}$Department of Theoretical Physics and Astrophysics, Faculty of Science, Masaryk University, Kotlarsk\'a 2, 611 37 Brno, Czech Republic\\
$^{6}$School of Science, Hiroshima University, 1-3-1 Kagamiyama, Higashi-Hiroshima 739-8526, Japan\\
$^{7}$KIPAC, Stanford University, 452 Lomita Mall, Stanford, CA 94305, USA\\
$^{8}$Department of Physics, Stanford University, 382 Via Pueblo Mall, Stanford, CA  94305-4060, USA\\
$^{9}$SLAC National Accelerator Laboratory, 2575 Sand Hill Road, Menlo Park, CA 94025, USA\\
$^{10}$ASTRON, Netherlands Institute for Radio Astronomy, PO Box 2, NL-7990 AA Dwingeloo, the Netherlands\\
$^{11}$Leiden Observatory, Leiden University, P.O. Box 9513, NL-2300 RA Leiden, the Netherlands\\
}}

\begin{document}
\maketitle

\begin{abstract}

We present the results from ALMA observations centred $40^{\prime\prime}$ (3 kpc in projection) southeast of the nucleus of M87. We report the detection of extended CO (2--1) line emission with a total flux of $(5.5 \pm 0.6) \times 10^{-18}$ erg s$^{-1}$ cm$^{-2}$ and corresponding molecular gas mass $M_{H_2}=(4.7 \pm 0.4) \times 10^5 M_\odot$, assuming a Galactic CO to H$_2$ conversion factor. ALMA data indicate a line-of-sight velocity of $-129\pm3$ km s$^{-1}$, in good agreement with measurements based on the [\ion{C}{ii}] and H$\alpha$+[\ion{N}{ii}] lines, and a velocity dispersion of $\sigma=27\pm3$ km s$^{-1}$. The CO(2--1) emission originates only outside the radio lobe of the AGN seen in the 6~cm VLA image, while the filament prolongs further inwards at other wavelengths. The molecular gas in M87 appears to be destroyed or excited by AGN activity, either by direct interaction with the radio plasma, or by the shock driven by the lobe into the X-ray emitting atmosphere. This is an important piece of the puzzle in understanding the impact of the central AGN on the amount of the coldest gas from which star formation can proceed. 

\end{abstract}

\begin{keywords}
galaxies: active Ð galaxies: clusters: individual (M87) Ð radio lines: galaxies
\end{keywords}

\section{Introduction}

`Cool-core' clusters of galaxies show sharp central X-ray surface brightness peaks and temperature dips. If the energy radiated away in the centers of these systems came only from the thermal energy of the hot, diffuse intra-cluster medium (ICM), the ICM would cool and form stars at rates orders of magnitude above what the observations suggest \citep[for a review, see][]{peterson2006}. The cooling is thought to be primarily offset by interactions between active galactic nuclei (AGN) in the central galaxies of these clusters and the surrounding ICM \citep[e.g.][]{McNamara2007}. 

Despite the ongoing AGN heating, line emission from cold atomic and molecular gas is frequently observed in cool cores \citep[e.g.][]{edge2001,salome2003}, and its distribution coincides with the lowest-temperature X-ray gas \citep[e.g.][]{sanders2007, werner2013}. This cold gas can either be intrinsic to the brightest central galaxy (having been supplied, perhaps, by a past merger with a gas-rich system or by stellar mass loss), or it can originate from the ICM \citep[e.g.][]{lim2008}. Even if cooling and heating are balanced on average throughout the volume of the cool core, thermal instabilities can occur when the local instantaneous ratio of the cooling time $t_c$ to the free-fall time $t_{ff}$ is low \citep{Sharma2012, gaspari2012}. Alternatively, the process of runaway cooling can be directly triggered by the AGN if filaments of low-temperature X-ray gas are dragged out of the cluster centres in the wakes of the buoyantly rising radio bubbles \citep{revaz2008,mcnamara2016}. The observed dynamics of the molecular gas clouds indeed indicate that this latter process may take place \citep{salome2011,russell2016,russell2017}. The cold gas is the reservoir from which star formation can proceed - therefore, in order to study the effect of AGN feedback on galaxy evolution, it is important to understand the relationship between AGN heating and the creation and/or destruction of the cold gas.

M~87 is the central dominant galaxy of the Virgo Cluster, the nearest cool-core cluster at a distance of only $\sim16.1$~Mpc \citep{blakeslee2009}. This allows us to view phenomena related to the interaction of the AGN jet inflated radio lobes with the surrounding environment in exceptional detail \citep[e.g.][]{forman2007,million2010b}. Deep XMM-Newton and Chandra observations have revealed the distribution of multi-phase plasma in the ICM, with the large-scale radio bubbles associated with the AGN having uplifted about 50\% of the low-temperature ($kT\sim1$ keV) gas from the cluster centre out to larger radii, effectively preventing it from cooling and forming stars \citep{simionescu2008a, werner2010}. A very small fraction of the 1~keV component is associated with even cooler X-ray gas, with $kT\sim0.5$~keV. In the core of the Perseus Cluster, another well-studied example of AGN feedback, the 0.5 keV gas phase traces an intricate network of optical H$\alpha$ and CO filaments \citep{sanders2007,salome2008}. In M~87, although there is a clear correlation between the 0.5 keV gas and H$\alpha$ emission \citep{werner2010}, the search for the associated molecular gas has so far yielded only upper limits \citep{salome2008} or, at best, very uncertain detections with low signal-to-noise \citep{tan2008}. More recently, {\it Herschel} Photoconductor Array Camera and Spectrometer (PACS) data allowed the discovery of [\ion{C}{ii}] emission from a prominent H$\alpha$ filamentary structure located south-east of the AGN \citep{werner2013}. This [\ion{C}{ii}] emission seems to be co-spatial not only with H$\alpha$, but also with far ultraviolet (FUV) filaments seen with the Solar Blind Channel (SBC) on the Advanced Camera for Surveys (ACS) of the Hubble Space Telescope (HST) \citep{sparks2009}. \citet{sparks2012} show that at least part of this UV flux originates from the \ion{C}{iv} line emitted by $10^5$ Kelvin gas. 

Here, we present new Atacama Large Millimeter Array (ALMA) observations of the coldest, molecular gas phase at the location of these filaments that were previously detected at other wavelengths.

\section{Observations and data analysis}
\label{analysis}

ALMA observed a region located $\sim$40\arcsec\ south-east of the AGN in M87 during Early Science Cycle 2, over 4 separate scheduling blocks executed between 2014 December 11 and 2015 January 3. The total observing time was 5.4 hours, of which 3 on source and 2.4 on calibrators. Ganymede and Callisto, along with Quasars 3C 273, J1058+0133, and J1229+0203 were used for amplitude, flux, and phase calibration. Between 34 and 41 of the 12m antennae were operational. The maximum baselines extended to 350 m, delivering a minimum angular resolution of 0.7\arcsec\ for a largest recoverable angular scale of $\sim$10\arcsec and a primary beam (field of view, FOV) of 28\arcsec. The J = 2 -- 1 rotational line transition of carbon monoxide at the redshift of M87 corresponds to a frequency of 229.5 GHz, which falls within the sensitivity range of ALMA's Band 6 receivers (211 -- 275 GHz). One baseband was centered on the CO(2--1) line, while the other three sampled the local continuum. 

The data were reduced using version 4.3.1 of the Common Astronomy Software Applications (CASA) software package \citep{mcmullin2007}. The continuum was subtracted by summing all line-free channels and removing the emission using the task \texttt{uvcontsub}. The final continuum-subtracted CO(2--1) data cube has an RMS noise of 0.2 mJy per beam per 50 km s$^{-1}$ channel. Measurement sets were imaged using Briggs weighting with a robust parameter of 0, yielding a beamsize of 1.4\arcsec $\times$ 0.7\arcsec\ at a position angle of 75 degrees. All ALMA images presented in this paper have been corrected for response of the primary beam. For consistency with the Herschel-PACS results \citep{werner2013}, we assume a systemic velocity of M87 of $v = 1307$ km s$^{-1}$, a redshift z=0.00436, and a luminosity distance $D_L=16.1$ Mpc \citep{blakeslee2009}, for which 1 arcmin corresponds to 4.65 kpc.

\section{Results}
\label{results}

\begin{figure}
\includegraphics[width=\columnwidth]{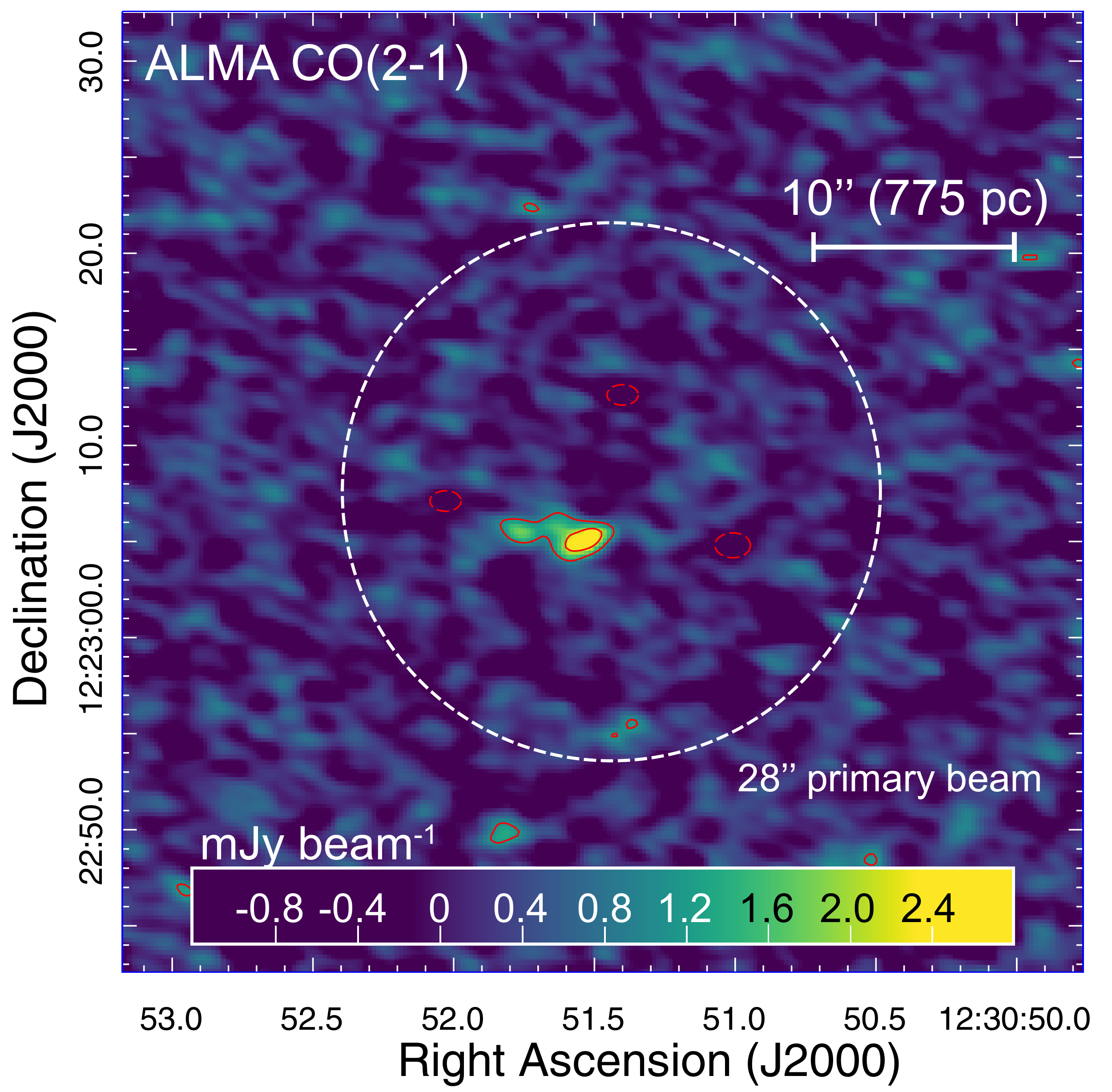}
\caption{Continuum-subtracted CO(2--1) ALMA image of the filamentary structure located south-east of the M87 nucleus. The CO(2--1) flux is integrated along the spectral axis from -300 km s$^{-1}$ to +300 km s$^{-1}$ relative to the galaxy's systemic velocity. 
The ALMA primary beam is indicated as a dashed white circle. Red contours indicate -3 (dashed), +3, and +6$\sigma$ levels.}
\label{raw}
\end{figure}

\begin{figure}
\centering
\includegraphics[width=0.9\columnwidth]{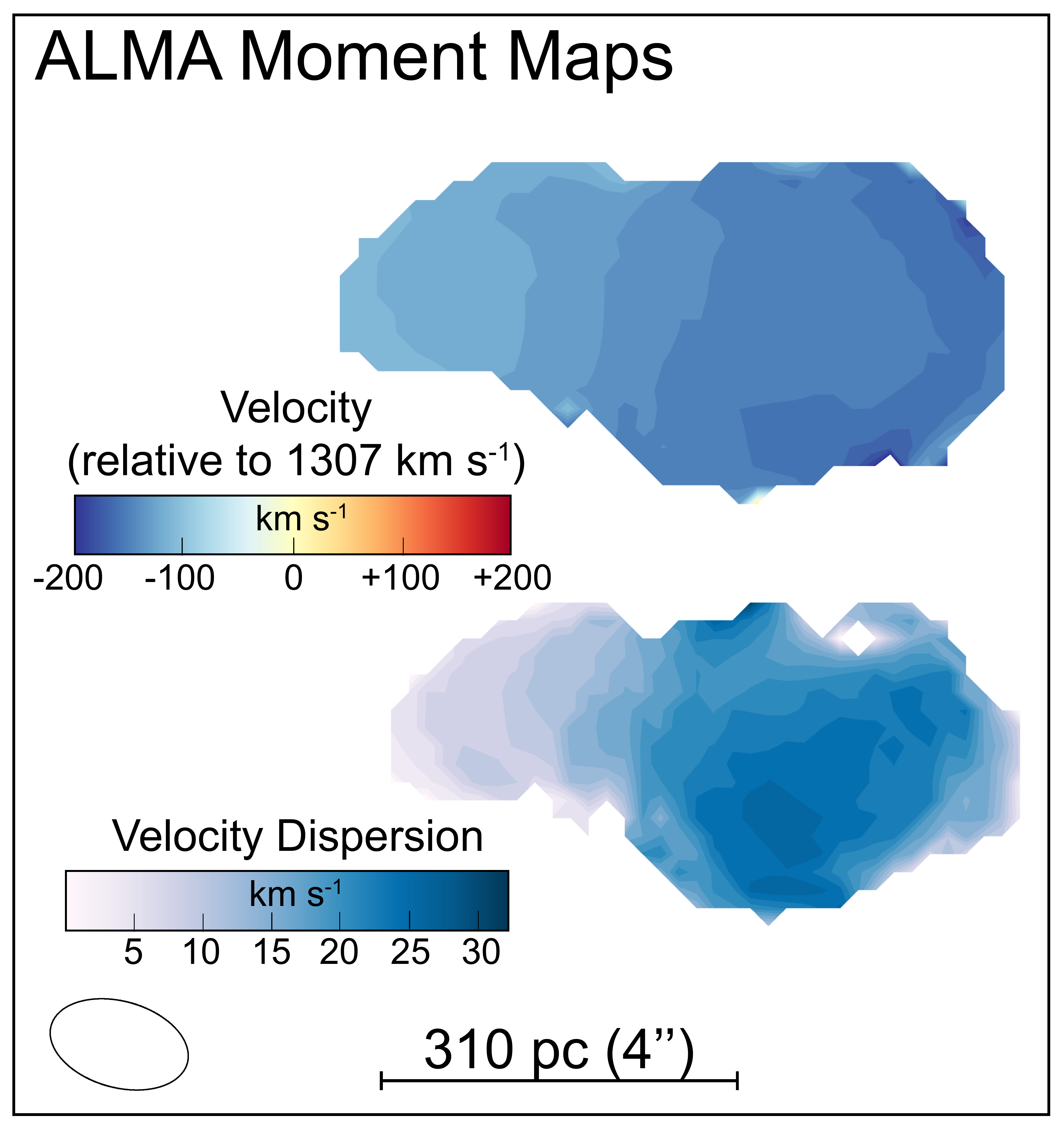}
\caption{The flux-weighted mean velocity and velocity dispersion of the molecular gas detected by ALMA, obtained using the ``masked-moment'' analysis. For clarity, we have zoomed in on the region where CO(2--1) emission is detected above the 3$\sigma$ threshold. The peak of the CO emission is located at $\alpha$ 12:30:51.54, $\delta$ 12:23:05.09.
The ALMA beam size is shown as a black ellipse.}
\label{dynamics}
\end{figure}

\begin{figure*}
\includegraphics[width=\textwidth]{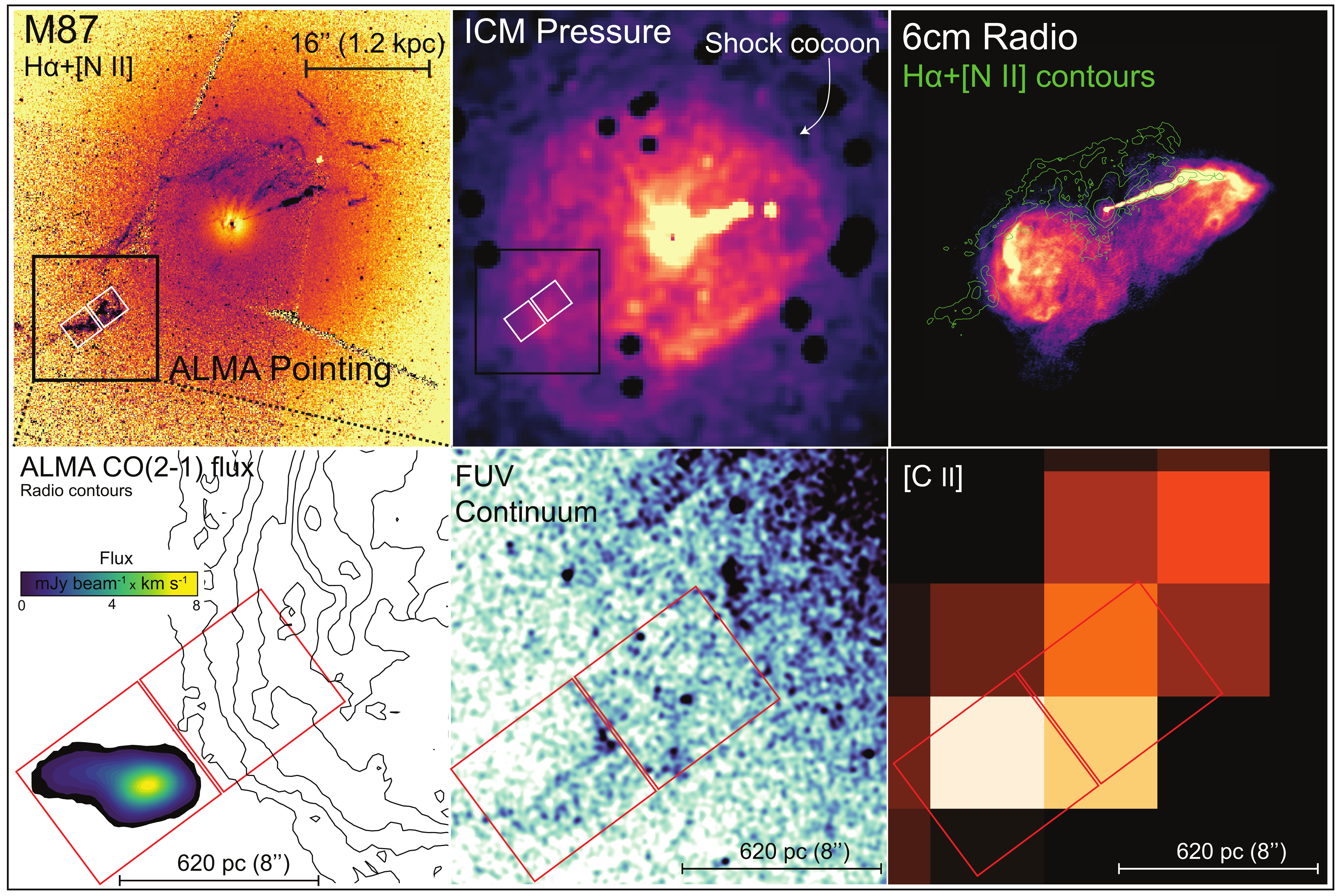}
\caption{Top panels show the H$\alpha$+[\ion{N}{ii}] flux from HST WFPC2 \citep{werner2013}, the Chandra X-ray pressure map \citep{million2010b}, and the 6~cm VLA radio emission \citep{hines1989}. The ALMA FOV is shown as a black box. All images in the top row share the same field of view.
The bottom row shows the ALMA CO(2--1) flux obtained from the masked moment analysis, FUV image from the HST ACS/SBC \citep{sparks2009}, and [\ion{C}{ii}] flux from Herschel PACS \citep{werner2013}.  
The $9\times9$ arcsec regions used for comparing the H$\alpha$ to CO(2--1) ratio across the edge of the radio lobe are shown in white (top left and top middle panels) or red (bottom panels). Contours of the 6~cm radio emission are shown in black in the ALMA panel, starting at 0.1 mJy and increasing towards the core in increments of a factor of two.}
\label{matrix}
\end{figure*}

The exquisite sensitivity of ALMA allows us to detect molecular gas in the nearest cool core cluster for the first time unambiguously, as shown in Figure \ref{raw}. The morphology of the detected CO emission consists of a brighter, approximately spherical clump, with a radius of about 150 pc (larger than the ALMA beam size), from which a narrower, fainter filament extends straight towards the east, over a projected distance of $\sim310$~pc. We have further employed the ``masked moment'' technique described by \citet{dame2011}\footnote{the code implementing this is publicly available from Timothy Davis at https://github.com/TimothyADavis/makeplots} to create a three-dimensional mask that takes into account coherence in position-velocity space to suppress the noise. Applying this analysis to a data cube binned to 10 km s$^{-1}$ channels, and keeping only the regions where the CO(2--1) line was detected with a significance above $3\sigma$, we obtain the CO(2--1) flux-weighted mean velocity and velocity dispersion shown in Figure \ref{dynamics}. The corresponding flux from the ``masked moment'' analysis (a smoothed version of the ``raw'' detection shown in Figure \ref{raw}) is reproduced in Figure \ref{matrix} and compared to detections of this filamentary structure at other wavelengths.

The total flux measured in the CO(2--1) line is $(5.5 \pm 0.6) \times 10^{-18}$ erg s$^{-1}$ cm$^{-2}$; the total molecular gas mass is then
\begin{equation*}
M_{\rm mol}=\frac{1.05\times10^4}{3.2}\left(\frac{X_{\rm CO}}{2\times10^{20} \frac{\rm cm^{-2}}{\rm K\: km\: s^{-1}}}\right)\left(\frac{1}{1+z}\right)\left(\frac{S_{\rm CO}\Delta\nu}{\rm Jy\:km\:s^{-1}}\right)\left(\frac{D_L}{\rm Mpc}\right)^2 M_\odot
\end{equation*}
where $S_{\rm CO}\Delta\nu = 0.56 \pm 0.06$ Jy km s$^{-1}$ is the emission integral. This mass estimate most critically relies on an assumption of the CO-to-H$_2$ conversion factor, $X_{\rm CO}$. Assuming the average Milky Way value of $X_{\rm CO} = 2\times10^{20} \frac{\rm cm^{-2}}{\rm K\: km\: s^{-1}}$ \citep{solomon1987} and a CO(2--1) to CO(1--0) flux density ratio of 3.2 \citep{david2014} yields a total molecular gas mass of $M_{\rm H_2}=(4.7 \pm 0.4) \times 10^5 M_\odot$. Note however that this value is subject to significant systematic uncertainty, as the Galactic conversion factor is not expected or observed to be universal \citep[for a review, see][]{bolatto2013}. 

The filamentary structure in the ALMA FOV has a line-of-sight (LOS) velocity of $v=-129\pm3$ km s$^{-1}$ and a narrow line width, $\sigma=27\pm3$ km s$^{-1}$. For comparison, \citet{sparks1993} report similar LOS velocities between -51 and -120 km s$^{-1}$ from H$\alpha$+[\ion{N}{ii}] observations (corrected to the systemic velocity assumed here), while [\ion{C}{ii}] measurements give $v=-123\pm5$ km s$^{-1}$ and $\sigma\sim55$ km s$^{-1}$ \citep[the latter at the limit of the Herschel PACS instrumental resolution;][]{werner2013}. The H$\alpha$+[\ion{N}{ii}] and [\ion{C}{ii}] emission extends at a fainter level towards the north and beyond the ALMA FOV, where it indicates receding velocities of $v\sim140$ km s$^{-1}$; no CO emission from this region is detected in our observation. 

Of further interest is the distribution of the CO flux with respect to the H$\alpha$+[\ion{N}{ii}] and FUV images, and the AGN radio lobe visible in the VLA observations (see Figure \ref{matrix}). The H$\alpha$+[\ion{N}{ii}] flux from the filamentary structure is roughly the same both inside and outside the 6~cm radio contours; the soft X-ray (0.7--0.9 keV) Chandra image is known to reveal a very similar filament morphology \citep{werner2010}. By contrast, CO emission is {\it only} observed outside of the radio lobe. \citet{million2010b} used deep Chandra observations of M87 to reconstruct the spatial distribution of the thermodynamic properties of the ambient ICM from spectroscopic mapping, and report the presence of a shock with a Mach number of at least 1.2, driven by the expansion of the 6~cm radio lobe; their pressure map is reproduced in the top middle panel of Figure \ref{matrix}, and shows that the CO emission is localised (in projection) outside of the 6~cm radio lobe but {\it inside} the shock cocoon propagating through the X-ray atmosphere of M87. 

To investigate this further, we consider two square regions, each with a size of $9\times9$\arcsec\, as shown in Figure \ref{matrix}. The inner region contains H$\alpha$+[\ion{N}{ii}] emission, but no CO emission, and is located fully inside the 6~cm radio lobe. The outer region contains both H$\alpha$+[\ion{N}{ii}] and CO(2--1) emission, and is located fully outside the 6~cm radio lobe. Both regions lie inside the shock detected in X-rays by \citet{million2010b}. Using the HST Wide Field and Planetary Camera 2 (WFPC2) F660N data, we obtain comparable H$\alpha$ fluxes for the two regions: $(3.1\pm0.3) \times10^{-14}$ and $(2.5\pm0.2) \times 10^{-14}$ erg s$^{-1}$ cm$^{-2}$ nearer and further from the AGN, respectively, where we have subtracted the expected contribution from [\ion{N}{ii}] and the galaxy stellar light. Meanwhile, the $3\sigma$ upper limit on the CO(2--1) flux from the inner region is $1.3\times10^{-18}$ erg s$^{-1}$ cm$^{-2}$, compared to $(5.5 \pm 0.6) \times 10^{-18}$ erg s$^{-1}$ cm$^{-2}$ detected in the outer region as reported above. Hence, the H$\alpha$ to CO(2--1) ratio changes by a factor of more than 5 across the edge of the radio lobe. Within the few ALMA spatial resolution elements where the CO flux is well determined, we find no evidence that the CO(2--1) to H$\alpha$ flux ratio varies along the filament. On the other hand, the [\ion{O}{iii}]/H$\beta$ emission line ratios reported by \citet{werner2013} show a strong decrease with radius, and thus have considerably smaller values in the regions associated with CO emission. This suggests that some form of interaction with the AGN depletes the reservoir of cold gas.
Compared to the expected correlation between the H$\alpha$ luminosity and molecular gas mass seen in other cluster central galaxies \citep[for the latest study, see][]{pulido2017}, M87 indeed appears very cold-gas poor given the H$\alpha$ flux measurements presented here. 


\section{Discussion}
\label{discussion}

Hints that AGN radio lobes may be responsible for heating the atomic/ionized filaments have been reported in several other cluster cores; for example, the high flux of the \ion{Fe}{x} coronal line, coincident with the edge of the radio lobe in the Centaurus Cluster \citep{canning2011}, or the unusually strong [\ion{O}{iii}] emission at the location where the AGN bubble/jet is interacting with the ionised gas in Abell 3581 \citep{canning2013}. 
\citet{morganti2016} show that the radio jet in Centaurus A also affects the atomic and ionized to molecular gas ratios, hinting that CO survives the jet-cloud interaction more easily than the other phases, which is opposite to the trend reported here.
Below, we speculate on possible scenarios that explain the interaction between the AGN and the molecular gas in M87, and comment briefly on the dynamics of the CO clouds.

\subsection{Interaction between the AGN and the molecular gas}

Since the boundary between the CO-emitting and CO-faint regions is located at the outer edge of the 6~cm emission, the first possible explanation is that the radio lobe interacts directly with the molecular gas, either depleting it or exciting it such that it primarily emits in higher order transitions (J = 3--2 or 4--3, rather than 2--1).  \citet{werner2013} show that internal shocks produced as the radio lobe pushes against the cold gas are likely responsible for the increased [\ion{O}{iii}]/H$\beta$ line ratios in the CO-faint regions of the filament, suggesting that an interaction is indeed taking place. If the filaments are threaded by magnetic fields, magnetic reconnection during the interaction with the magnetised radio lobes may further contribute to their destruction \citep[some aspects of heating of the cold gas filaments by reconnection are discussed in ][]{churazov2013}. 

\begin{figure}
\begin{center}
\includegraphics[width=0.9\columnwidth]{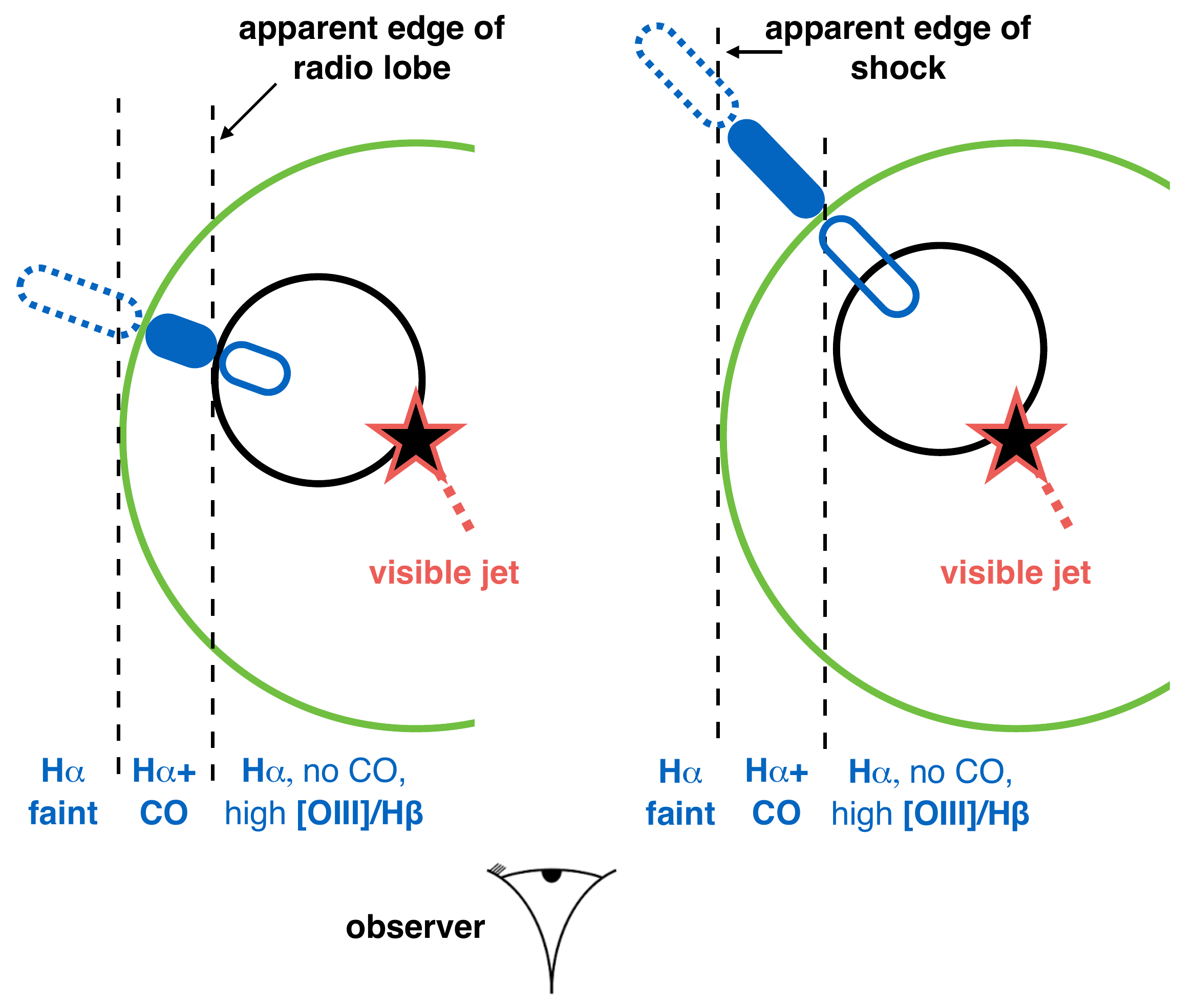}
\end{center}

\caption{Illustration of the effects of LOS projection. The observed filaments are sketched in blue, the X-ray shock in green, and the radio lobe in black. If the inclination is too high, the change in H$\alpha$+[\ion{N}{ii}] brightness cannot be due to the shock and would remain unexplained.
}\label{schematic}
\end{figure}

Another possibility is that the molecular gas is destroyed or heated by the passage of the X-ray shock, rather than the interaction with the radio lobe. In principle, for a large inclination of the molecular filaments with respect to the plane of the sky, the boundary between the CO-emitting and CO-faint regions may lie at the position of the X-ray shock, while appearing to coincide with the edge of the radio lobe in projection (see right panel of Figure \ref{schematic}). HST observations of the superluminal motion in the jet of M87 \citep{biretta1999}, as well as the geometry and radio polarisation of the large-scale radio lobes seen in the 90 cm radio image, do in fact suggest that the orientation of the AGN jet and lobes is close to the line of sight \citep[for a detailed discussion, see][]{werner2010}. 

However, the position of the X-ray shock corresponds in projection to a marked drop in the H$\alpha$+[\ion{N}{ii}] luminosity of the filamentary nebula \citep{werner2010}. The likely explanation is that the propagation of the shock introduces shearing at the interface between the hot and cold gas, enabling ICM particles to penetrate the filaments more easily and dissociate/ionize the atomic gas, and thereby powering the H$\alpha$+[\ion{N}{ii}] emission \citep{ferland2009,werner2010}. 
This scenario requires the optical line emitting filaments to be oriented relatively close to the plane of the sky; a significant inclination would imply either that the sharp drop in H$\alpha$+[\ion{N}{ii}] luminosity should be seen at smaller projected radii than the edge of the shock, or that the change in H$\alpha$+[\ion{N}{ii}] brightness is not due to the shock and would thus remain unexplained.

We conclude that it is unlikely that the X-ray shock coincides in 3D geometry with the dimming of CO(2--1) emission observed with ALMA. It is nevertheless possible that the X-ray shock is still responsible for destroying or heating the molecular gas, but that there is a time lag between the passage of the shock and the observed dimming of the CO emission. For example, the turbulence and mixing induced by the passage of the shock could affect the atomic/ionized gas more quickly, but may require a longer time to fully destroy the CO lumps. In this case, the inner molecular gas has been depleted, while that located closer to the shock front is still in the process of being broken up. The fact that the molecular mass to H$\alpha$ luminosity ratio is very low compared to the correlation seen in other cluster central galaxies \citep{salome2003,pulido2017}, even in the region where CO is detected, seems to support this interpretation. The distance between the outer edge of the X-ray shock and the inner edge of the detected CO gas is 9 arcsec, corresponding to 0.7 kpc. For a Mach number of the X-ray shock of 1.2 \citep{million2010b}, a sound speed of 625 km s$^{-1}$ corresponding to an assumed ambient temperature of 1.5 keV, and given an orientation of the filaments close to the plane of the sky as argued above, we estimate that in this scenario 1.1 Myr have elapsed between the shock passage and the observed destruction of the molecular gas. While conserving the Mach number, the shock propagation through the cold gas with low sound speeds should also slow down significantly, which could further contribute to the difference between the present location of the X-ray shock, brightened H$\alpha$+[\ion{N}{ii}] filament, and the CO-depleted area. In this case, the association between the edge of the radio lobe and the inner edge of the CO cloud would have to be accidental. 


We further note that, according to estimates by \citet{sparks2009}, a total mass of cold gas converted into stars of only $\sim2000 M_\odot$ should result in 14 individual young stars detected in the HST FUV image. It is therefore unlikely that the interaction with the AGN stimulated the collapse of the molecular gas into stars, rather than heating it directly. Our observations thus show that AGN feedback, either through shocks in the surrounding ICM or the relativistic radio lobes, has an important impact on the amount and emissivity of the coldest gas from which star formation can proceed. While some earlier ALMA observations have suggested that the AGN can stimulate the production of cold gas by cooling instabilities in the wakes of rising radio bubbles \citep{russell2016,russell2017}, here we show that AGN activity may also have the opposite effect in destroying already existing molecular gas.

\subsection{Dynamics of the molecular gas in M87}

\citet{werner2013} report a LOS velocity gradient of about 300 km s$^{-1}$ along the [\ion{C}{ii}] filamentary structure observed with Herschel; while we can only observe a small fraction of this filament in CO with ALMA, the measured velocity dispersion appears remarkably quiescent ($\sigma=27\pm3$ km s$^{-1}$), an order of magnitude smaller than the total line-of-sight velocity gradient. However, this value is still higher than that of resolved giant molecular clouds in the Milky Way, where line widths seldom exceed 10 km s$^{-1}$ \citep{solomon1987}. As discussed previously for other observations of cluster cores \citep[see e.g.][]{david2014}, it is unlikely therefore that this structure consists of a single cloud; rather, it is probably a giant molecular association composed of several cloudlets.
In this case, we would expect the parts of the filament that appear brighter in CO to represent the superposition of a larger number of cloudlets and hence sample a larger velocity dispersion. This is consistent with the trend seen in Figure \ref{dynamics}, which shows a gradient in $\sigma$ from values as low as 5--10 km s$^{-1}$ at the faintest, eastern end of the filament up to about 30 km s$^{-1}$ near the peak of the CO flux. We note that, for a velocity dispersion $\sigma\sim30$ km s$^{-1}$ and a linear extent $R_{\rm cl}$ of a few hundred pc, the total mass required to keep the molecular gas cloudlets gravitationally self-bound, $M_{vir}\sim\sigma^2R_{\rm cl}/G$ with G the gravitational constant, is of order a few times $10^7 M_\odot$. This is nearly two orders of magnitude larger than the mass estimated from the CO flux. Hence, the observed structure would have to be very short-lived, unless its dynamics and evolution are influenced by an external factor such the surrounding ICM.

Recent ALMA observations of the Phoenix cluster revealed a remarkable similarity in the molecular gas velocities at the outer tips of all the observed filaments, separated by $\sim$30 kpc, hinting that this remote molecular gas could be coupled to the hot atmosphere \citep{russell2017}. So far the only direct measurements of the velocities in the X-ray emitting ICM were performed for the Perseus Cluster \citep{hitomi2016}, and show good agreement with the velocity gradients and velocity dispersion inferred for the optical H$\alpha$-emitting nebula \citep{hatch2006} and molecular CO \citep{salome2011}.

If the dynamics of molecular cloudlets is coupled to the ambient X-ray gas, the remarkably low velocity dispersion measured by ALMA suggests that the ICM in the core of M87, despite signs of vigorous AGN activity, has low turbulent velocities, and that the motions are predominantly laminar. This is consistent with the conclusions of \citet{zhuravleva2014}, who used the power spectrum of surface brightness fluctuations from deep Chandra observations to predict turbulent velocities in the core of M87 of $\sigma\sim$60 km s$^{-1}$, though on slightly different spatial scales than the current ALMA observation. This is encouraging for using both Chandra surface brightness fluctuations and ALMA and optical spectra to predict the ICM dynamics that will be measured through future high-resolution X-ray spectroscopy with \textit{Athena} or the X-ray Astronomy Recovery Mission (XARM).

\section{Conclusions}
\label{conclusions}

ALMA Band 6 observations centred $40^{\prime\prime}$ (3 kpc in projection) southeast of the nucleus of M87 allowed us to measure a total flux of $(5.5 \pm 0.6) \times 10^{-18}$ erg s$^{-1}$ cm$^{-2}$ in the  J = 2 -- 1 rotational line transition of carbon monoxide, representing the first secure detection of molecular gas in the brightest cluster galaxy of the nearest cool-core cluster. This flux translates to a total molecular gas mass of $M_{\rm H_2}=(4.7 \pm 0.4) \times 10^5 M_\odot$, albeit with large uncertainties related to the assumption that the CO-to-H$_2$ conversion factor is similar to that in our own Galaxy. The molecular gas detected with ALMA extends over a total length of about 600~pc (elongated in the east--west direction) and is co-spatial with a multi-phase filament previously identified at several wavelengths, from soft X-rays with Chandra to H$\alpha$+[\ion{N}{ii}] and FUV with HST, and [\ion{C}{ii}] with Herschel PACS. 
The ratio of molecular gas mass to H$\alpha$ luminosity in M87 lies well below typical values found in other cluster central galaxies with reported CO detections.

The distribution of CO(2--1) flux with respect to H$\alpha$+[\ion{N}{ii}] shows an intriguing radial trend. All of the CO emission appears to be located outside the radio lobe of the AGN seen in the 6~cm VLA image, while the H$\alpha$+[\ion{N}{ii}] and soft X-ray fluxes are similar both inside and outside the radio lobe. Using the $3\sigma$ confidence upper limit from ALMA for the region located inside the radio lobe implies a change in the H$\alpha$ to CO(2--1) ratio of at least a factor of 5. It appears that either relativistic particles and magnetic fields from this radio lobe, or the shock that this lobe is known to drive into the X-ray emitting atmosphere, are able to destroy the molecular gas in M87, or excite it such that it primarily emits in higher order (CO J = 3--2 or 4--3) transitions. This is an important piece of the puzzle in quantifying the impact of the central AGN on the amount of the coldest gas from which star formation can proceed. 

The molecular gas observed with ALMA has a small velocity dispersion, $\sigma=27\pm3$ km s$^{-1}$. This is an order of magnitude smaller than the total line-of-sight velocity gradient observed along the larger-scale [\ion{C}{ii}] filaments that the CO gas is embedded in, and is consistent with a predominantly laminar dynamical structure of these filaments, as already suggested in other cool-core clusters.

\section*{Acknowledgments}

We thank B. R. McNamara and an anonymous referee for their suggestions and comments. This paper makes use of the following ALMA data: ADS/JAO.ALMA\#2013.1.00862.S. ALMA is a partnership of ESO (representing its member states), NSF (USA) and NINS (Japan), together with NRC (Canada), MOST and ASIAA (Taiwan), and KASI (Republic of Korea), in cooperation with the Republic of Chile. The Joint ALMA Observatory is operated by ESO, AUI/NRAO and NAOJ. This work was supported in part by the Lend\"ulet LP2016-11 grant awarded by the Hungarian Academy of Sciences and by the US Department of Energy under contract number DE-AC02-76SF00515.

\bibliographystyle{mnras}
\bibliography{clusters_alma}

\begin{thebibliography}{}
\makeatletter
\relax
\def\mn@urlcharsother{\let\do\@makeother \do\$\do\&\do\#\do\^\do\_\do\%\do\~}
\def\mn@doi{\begingroup\mn@urlcharsother \@ifnextchar [ {\mn@doi@}
  {\mn@doi@[]}}
\def\mn@doi@[#1]#2{\def\@tempa{#1}\ifx\@tempa\@empty \href
  {http://dx.doi.org/#2} {doi:#2}\else \href {http://dx.doi.org/#2} {#1}\fi
  \endgroup}
\def\mn@eprint#1#2{\mn@eprint@#1:#2::\@nil}
\def\mn@eprint@arXiv#1{\href {http://arxiv.org/abs/#1} {{\tt arXiv:#1}}}
\def\mn@eprint@dblp#1{\href {http://dblp.uni-trier.de/rec/bibtex/#1.xml}
  {dblp:#1}}
\def\mn@eprint@#1:#2:#3:#4\@nil{\def\@tempa {#1}\def\@tempb {#2}\def\@tempc
  {#3}\ifx \@tempc \@empty \let \@tempc \@tempb \let \@tempb \@tempa \fi \ifx
  \@tempb \@empty \def\@tempb {arXiv}\fi \@ifundefined
  {mn@eprint@\@tempb}{\@tempb:\@tempc}{\expandafter \expandafter \csname
  mn@eprint@\@tempb\endcsname \expandafter{\@tempc}}}

\bibitem[\protect\citeauthoryear{{Biretta}, {Sparks}  \& {Macchetto}}{{Biretta}
  et~al.}{1999}]{biretta1999}
{Biretta} J.~A.,  {Sparks} W.~B.,   {Macchetto} F.,  1999, \mn@doi [\apj]
  {10.1086/307499}, \href {http://adsabs.harvard.edu/abs/1999ApJ...520..621B}
  {520, 621}

\bibitem[\protect\citeauthoryear{{Blakeslee} et~al.,}{{Blakeslee}
  et~al.}{2009}]{blakeslee2009}
{Blakeslee} J.~P.,  et~al., 2009, \mn@doi [\apj] {10.1088/0004-637X/694/1/556},
  \href {http://adsabs.harvard.edu/abs/2009ApJ...694..556B} {694, 556}

\bibitem[\protect\citeauthoryear{{Bolatto}, {Wolfire}  \& {Leroy}}{{Bolatto}
  et~al.}{2013}]{bolatto2013}
{Bolatto} A.~D.,  {Wolfire} M.,   {Leroy} A.~K.,  2013, \mn@doi [\araa]
  {10.1146/annurev-astro-082812-140944}, \href
  {http://adsabs.harvard.edu/abs/2013ARA%26A..51..207B} {51, 207}

\bibitem[\protect\citeauthoryear{{Canning}, {Fabian}, {Johnstone}, {Sanders},
  {Crawford}, {Ferland}  \& {Hatch}}{{Canning} et~al.}{2011}]{canning2011}
{Canning} R.~E.~A.,  {Fabian} A.~C.,  {Johnstone} R.~M.,  {Sanders} J.~S.,
  {Crawford} C.~S.,  {Ferland} G.~J.,   {Hatch} N.~A.,  2011, \mn@doi [\mnras]
  {10.1111/j.1365-2966.2011.19470.x}, \href
  {http://adsabs.harvard.edu/abs/2011MNRAS.417.3080C} {417, 3080}

\bibitem[\protect\citeauthoryear{{Canning} et~al.,}{{Canning}
  et~al.}{2013}]{canning2013}
{Canning} R.~E.~A.,  et~al., 2013, \mn@doi [\mnras] {10.1093/mnras/stt1345},
  \href {http://adsabs.harvard.edu/abs/2013MNRAS.435.1108C} {435, 1108}

\bibitem[\protect\citeauthoryear{{Churazov}, {Ruszkowski}  \&
  {Schekochihin}}{{Churazov} et~al.}{2013}]{churazov2013}
{Churazov} E.,  {Ruszkowski} M.,   {Schekochihin} A.,  2013, \mn@doi [\mnras]
  {10.1093/mnras/stt1594}, \href
  {http://adsabs.harvard.edu/abs/2013MNRAS.436..526C} {436, 526}

\bibitem[\protect\citeauthoryear{{Dame}}{{Dame}}{2011}]{dame2011}
{Dame} T.~M.,  2011, preprint, \href
  {http://adsabs.harvard.edu/abs/2011arXiv1101.1499D} {} (\mn@eprint {arXiv}
  {1101.1499})

\bibitem[\protect\citeauthoryear{{David} et~al.,}{{David}
  et~al.}{2014}]{david2014}
{David} L.~P.,  et~al., 2014, \mn@doi [\apj] {10.1088/0004-637X/792/2/94},
  \href {http://adsabs.harvard.edu/abs/2014ApJ...792...94D} {792, 94}

\bibitem[\protect\citeauthoryear{{Edge}}{{Edge}}{2001}]{edge2001}
{Edge} A.~C.,  2001, \mnras, \href
  {http://adsabs.harvard.edu/cgi-bin/nph-bib_query?bibcode=2001MNRAS.328..762E&db_key=AST}
  {328, 762}

\bibitem[\protect\citeauthoryear{{Ferland}, {Fabian}, {Hatch}, {Johnstone},
  {Porter}, {van Hoof}  \& {Williams}}{{Ferland} et~al.}{2009}]{ferland2009}
{Ferland} G.~J.,  {Fabian} A.~C.,  {Hatch} N.~A.,  {Johnstone} R.~M.,  {Porter}
  R.~L.,  {van Hoof} P.~A.~M.,   {Williams} R.~J.~R.,  2009, \mn@doi [\mnras]
  {10.1111/j.1365-2966.2008.14153.x}, \href
  {http://cdsads.u-strasbg.fr/abs/2009MNRAS.392.1475F} {392, 1475}

\bibitem[\protect\citeauthoryear{{Forman} et~al.,}{{Forman}
  et~al.}{2007}]{forman2007}
{Forman} W.,  et~al., 2007, \mn@doi [\apj] {10.1086/519480}, \href
  {http://cdsads.u-strasbg.fr/abs/2007ApJ...665.1057F} {665, 1057}

\bibitem[\protect\citeauthoryear{{Gaspari}, {Ruszkowski}  \&
  {Sharma}}{{Gaspari} et~al.}{2012}]{gaspari2012}
{Gaspari} M.,  {Ruszkowski} M.,   {Sharma} P.,  2012, \mn@doi [\apj]
  {10.1088/0004-637X/746/1/94}, \href
  {http://adsabs.harvard.edu/abs/2012ApJ...746...94G} {746, 94}

\bibitem[\protect\citeauthoryear{{Hatch}, {Crawford}, {Johnstone}  \&
  {Fabian}}{{Hatch} et~al.}{2006}]{hatch2006}
{Hatch} N.~A.,  {Crawford} C.~S.,  {Johnstone} R.~M.,   {Fabian} A.~C.,  2006,
  \mn@doi [\mnras] {10.1111/j.1365-2966.2006.09970.x}, \href
  {http://cdsads.u-strasbg.fr/abs/2006MNRAS.367..433H} {367, 433}

\bibitem[\protect\citeauthoryear{{Hines}, {Eilek}  \& {Owen}}{{Hines}
  et~al.}{1989}]{hines1989}
{Hines} D.~C.,  {Eilek} J.~A.,   {Owen} F.~N.,  1989, \mn@doi [\apj]
  {10.1086/168163}, \href {http://cdsads.u-strasbg.fr/abs/1989ApJ...347..713H}
  {347, 713}

\bibitem[\protect\citeauthoryear{{Hitomi Collaboration}}{{Hitomi
  Collaboration}}{2016}]{hitomi2016}
{Hitomi Collaboration} 2016, \mn@doi [\nat] {10.1038/nature18627}, \href
  {http://adsabs.harvard.edu/abs/2016Natur.535..117H} {535, 117}

\bibitem[\protect\citeauthoryear{{Lim}, {Ao}  \& {Dinh-V-Trung}}{{Lim}
  et~al.}{2008}]{lim2008}
{Lim} J.,  {Ao} Y.,   {Dinh-V-Trung} 2008, \mn@doi [\apj] {10.1086/523664},
  \href {http://adsabs.harvard.edu/abs/2008ApJ...672..252L} {672, 252}

\bibitem[\protect\citeauthoryear{{McMullin}, {Waters}, {Schiebel}, {Young}  \&
  {Golap}}{{McMullin} et~al.}{2007}]{mcmullin2007}
{McMullin} J.~P.,  {Waters} B.,  {Schiebel} D.,  {Young} W.,   {Golap} K.,
  2007, in {Shaw} R.~A.,  {Hill} F.,   {Bell} D.~J.,  eds,  Astronomical
  Society of the Pacific Conference Series Vol. 376, Astronomical Data Analysis
  Software and Systems XVI. p.~127

\bibitem[\protect\citeauthoryear{{McNamara} \& {Nulsen}}{{McNamara} \&
  {Nulsen}}{2007}]{McNamara2007}
{McNamara} B.~R.,  {Nulsen} P.~E.~J.,  2007, \mn@doi [ARA\&A]
  {10.1146/annurev.astro.45.051806.110625}, \href
  {http://cdsads.u-strasbg.fr/abs/2007ARA%26A..45..117M} {45, 117}

\bibitem[\protect\citeauthoryear{{McNamara}, {Russell}, {Nulsen}, {Hogan},
  {Fabian}, {Pulido}  \& {Edge}}{{McNamara} et~al.}{2016}]{mcnamara2016}
{McNamara} B.~R.,  {Russell} H.~R.,  {Nulsen} P.~E.~J.,  {Hogan} M.~T.,
  {Fabian} A.~C.,  {Pulido} F.,   {Edge} A.~C.,  2016, \mn@doi [\apj]
  {10.3847/0004-637X/830/2/79}, \href
  {http://adsabs.harvard.edu/abs/2016ApJ...830...79M} {830, 79}

\bibitem[\protect\citeauthoryear{{Million}, {Werner}, {Simionescu}, {Allen},
  {Nulsen}, {Fabian}, {B{\"o}hringer}  \& {Sanders}}{{Million}
  et~al.}{2010}]{million2010b}
{Million} E.~T.,  {Werner} N.,  {Simionescu} A.,  {Allen} S.~W.,  {Nulsen}
  P.~E.~J.,  {Fabian} A.~C.,  {B{\"o}hringer} H.,   {Sanders} J.~S.,  2010,
  \mn@doi [\mnras] {10.1111/j.1365-2966.2010.17220.x}, \href
  {http://adsabs.harvard.edu/abs/2010MNRAS.407.2046M} {407, 2046}

\bibitem[\protect\citeauthoryear{{Morganti}, {Oosterloo}, {Oonk}, {Santoro}  \&
  {Tadhunter}}{{Morganti} et~al.}{2016}]{morganti2016}
{Morganti} R.,  {Oosterloo} T.,  {Oonk} J.~B.~R.,  {Santoro} F.,   {Tadhunter}
  C.,  2016, \mn@doi [\aap] {10.1051/0004-6361/201628950}, \href
  {http://adsabs.harvard.edu/abs/2016A%26A...592L...9M} {592, L9}

\bibitem[\protect\citeauthoryear{{Peterson} \& {Fabian}}{{Peterson} \&
  {Fabian}}{2006}]{peterson2006}
{Peterson} J.~R.,  {Fabian} A.~C.,  2006, \mn@doi [Phys.~Rep.]
  {10.1016/j.physrep.2005.12.007}, \href
  {http://adsabs.harvard.edu/abs/2006PhR...427....1P} {427, 1}

\bibitem[\protect\citeauthoryear{{Pulido} et~al.,}{{Pulido}
  et~al.}{2017}]{pulido2017}
{Pulido} F.~A.,  et~al., 2017, preprint, \href
  {http://adsabs.harvard.edu/abs/2017arXiv171004664P} {} (\mn@eprint {arXiv}
  {1710.04664})

\bibitem[\protect\citeauthoryear{{Revaz}, {Combes}  \& {Salom{\'e}}}{{Revaz}
  et~al.}{2008}]{revaz2008}
{Revaz} Y.,  {Combes} F.,   {Salom{\'e}} P.,  2008, \mn@doi [\aap]
  {10.1051/0004-6361:20078915}, \href
  {http://adsabs.harvard.edu/abs/2008A%26A...477L..33R} {477, L33}

\bibitem[\protect\citeauthoryear{{Russell} et~al.,}{{Russell}
  et~al.}{2016}]{russell2016}
{Russell} H.~R.,  et~al., 2016, \mn@doi [\mnras] {10.1093/mnras/stw409}, \href
  {http://adsabs.harvard.edu/abs/2016MNRAS.458.3134R} {458, 3134}

\bibitem[\protect\citeauthoryear{{Russell} et~al.,}{{Russell}
  et~al.}{2017}]{russell2017}
{Russell} H.~R.,  et~al., 2017, \mn@doi [\apj] {10.3847/1538-4357/836/1/130},
  \href {http://adsabs.harvard.edu/abs/2017ApJ...836..130R} {836, 130}

\bibitem[\protect\citeauthoryear{{Salom{\'e}} \& {Combes}}{{Salom{\'e}} \&
  {Combes}}{2003}]{salome2003}
{Salom{\'e}} P.,  {Combes} F.,  2003, \mn@doi [\aap]
  {10.1051/0004-6361:20031438}, \href
  {http://adsabs.harvard.edu/abs/2003A%26A...412..657S} {412, 657}

\bibitem[\protect\citeauthoryear{{Salom{\'e}} \& {Combes}}{{Salom{\'e}} \&
  {Combes}}{2008}]{salome2008}
{Salom{\'e}} P.,  {Combes} F.,  2008, \mn@doi [\aap]
  {10.1051/0004-6361:200810262}, \href
  {http://cdsads.u-strasbg.fr/abs/2008A%26A...489..101S} {489, 101}

\bibitem[\protect\citeauthoryear{{Salom{\'e}}, {Combes}, {Revaz}, {Downes},
  {Edge}  \& {Fabian}}{{Salom{\'e}} et~al.}{2011}]{salome2011}
{Salom{\'e}} P.,  {Combes} F.,  {Revaz} Y.,  {Downes} D.,  {Edge} A.~C.,
  {Fabian} A.~C.,  2011, \mn@doi [\aap] {10.1051/0004-6361/200811333}, \href
  {http://adsabs.harvard.edu/abs/2011A%26A...531A..85S} {531, A85}

\bibitem[\protect\citeauthoryear{{Sanders} \& {Fabian}}{{Sanders} \&
  {Fabian}}{2007}]{sanders2007}
{Sanders} J.~S.,  {Fabian} A.~C.,  2007, \mn@doi [\mnras]
  {10.1111/j.1365-2966.2007.12347.x}, \href
  {http://cdsads.u-strasbg.fr/abs/2007MNRAS.381.1381S} {381, 1381}

\bibitem[\protect\citeauthoryear{{Sharma}, {McCourt}, {Quataert}  \&
  {Parrish}}{{Sharma} et~al.}{2012}]{Sharma2012}
{Sharma} P.,  {McCourt} M.,  {Quataert} E.,   {Parrish} I.~J.,  2012, \mn@doi
  [\mnras] {10.1111/j.1365-2966.2011.20246.x}, \href
  {http://adsabs.harvard.edu/abs/2012MNRAS.420.3174S} {420, 3174}

\bibitem[\protect\citeauthoryear{{Simionescu}, {Werner}, {Finoguenov},
  {B{\"o}hringer}  \& {Br{\"u}ggen}}{{Simionescu}
  et~al.}{2008}]{simionescu2008a}
{Simionescu} A.,  {Werner} N.,  {Finoguenov} A.,  {B{\"o}hringer} H.,
  {Br{\"u}ggen} M.,  2008, \mn@doi [\aap] {10.1051/0004-6361:20078749}, \href
  {http://cdsads.u-strasbg.fr/abs/2008A%26A...482...97S} {482, 97}

\bibitem[\protect\citeauthoryear{{Solomon}, {Rivolo}, {Barrett}  \&
  {Yahil}}{{Solomon} et~al.}{1987}]{solomon1987}
{Solomon} P.~M.,  {Rivolo} A.~R.,  {Barrett} J.,   {Yahil} A.,  1987, \mn@doi
  [\apj] {10.1086/165493}, \href
  {http://adsabs.harvard.edu/abs/1987ApJ...319..730S} {319, 730}

\bibitem[\protect\citeauthoryear{{Sparks}, {Ford}  \& {Kinney}}{{Sparks}
  et~al.}{1993}]{sparks1993}
{Sparks} W.~B.,  {Ford} H.~C.,   {Kinney} A.~L.,  1993, \mn@doi [\apj]
  {10.1086/173022}, \href {http://cdsads.u-strasbg.fr/abs/1993ApJ...413..531S}
  {413, 531}

\bibitem[\protect\citeauthoryear{{Sparks}, {Pringle}, {Donahue}, {Carswell},
  {Voit}, {Cracraft}  \& {Martin}}{{Sparks} et~al.}{2009}]{sparks2009}
{Sparks} W.~B.,  {Pringle} J.~E.,  {Donahue} M.,  {Carswell} R.,  {Voit} M.,
  {Cracraft} M.,   {Martin} R.~G.,  2009, \mn@doi [\apjl]
  {10.1088/0004-637X/704/1/L20}, \href
  {http://adsabs.harvard.edu/abs/2009ApJ...704L..20S} {704, L20}

\bibitem[\protect\citeauthoryear{{Sparks} et~al.,}{{Sparks}
  et~al.}{2012}]{sparks2012}
{Sparks} W.~B.,  et~al., 2012, \mn@doi [\apjl] {10.1088/2041-8205/750/1/L5},
  \href {http://adsabs.harvard.edu/abs/2012ApJ...750L...5S} {750, L5}

\bibitem[\protect\citeauthoryear{{Tan}, {Beuther}, {Walter}  \&
  {Blackman}}{{Tan} et~al.}{2008}]{tan2008}
{Tan} J.~C.,  {Beuther} H.,  {Walter} F.,   {Blackman} E.~G.,  2008, \mn@doi
  [\apj] {10.1086/592592}, \href
  {http://adsabs.harvard.edu/abs/2008ApJ...689..775T} {689, 775}

\bibitem[\protect\citeauthoryear{{Werner} et~al.,}{{Werner}
  et~al.}{2010}]{werner2010}
{Werner} N.,  et~al., 2010, \mn@doi [\mnras]
  {10.1111/j.1365-2966.2010.16755.x}, \href
  {http://adsabs.harvard.edu/abs/2010MNRAS.407.2063W} {407, 2063}

\bibitem[\protect\citeauthoryear{{Werner} et~al.,}{{Werner}
  et~al.}{2013}]{werner2013}
{Werner} N.,  et~al., 2013, \mn@doi [\apj] {10.1088/0004-637X/767/2/153}, \href
  {http://adsabs.harvard.edu/abs/2013ApJ...767..153W} {767, 153}

\bibitem[\protect\citeauthoryear{{Zhuravleva} et~al.,}{{Zhuravleva}
  et~al.}{2014}]{zhuravleva2014}
{Zhuravleva} I.,  et~al., 2014, \mn@doi [\nat] {10.1038/nature13830}, \href
  {http://adsabs.harvard.edu/abs/2014Natur.515...85Z} {515, 85}

\makeatother
\end{thebibliography}

\end{document}